\documentstyle[12pt]{article}
\title{An Algorithm for the Longest Common Subsequence and Substring Problem for Multiple Strings}
\author{Rao Li \\
 			Dept. of Computer Science, Engineering, and Mathematics \\
        University of South Carolina Aiken \\
	      Aiken, SC 29801 \\
        USA \\
       {\it Email: raol@usca.edu }
       }
\date{Nov. 23, 2024}

\begin{document}
\maketitle
\begin{abstract}
Let $X_1, X_2, ..., X_s$ and $Y_1, Y_2, ..., Y_t$ be strings over an alphabet $\Sigma$, where $s$ and $t$ are positive integers. 
The longest common subsequence and substring problem for multiple strings $X_1, X_2, ..., X_s$ and $Y_1, Y_2, ..., Y_t$ is to find the longest string which is a subsequence of $X_1, X_2, ..., X_s$ and a substring of $Y_1, Y_2, ..., Y_t$. In this paper, 
we propose an algorithm to solve the problem.    
 \end{abstract} 
$$Keywords: Algorithm, the \,\, longest \,\, common \,\,  subsequence \,\, and \,\, substring$$  
  $$problem, \,\, the \,\, longest \,\, common \,\,  subsequence \,\, and \,\, substring \,\, problem \,\, for$$ 
$$multiple \,\, strings.$$ 
$$AMS \,\, Mathematics \,\, Subject \,\, Classification (2020): 68W32, 68W40$$
  \newpage

\noindent {\bf 1.  Introduction} \\

Let $\Sigma$ be an alphabet and $S$ a string over $\Sigma$. A subsequence of a string $S$ is obtained by deleting zero or more elements of $S$.
A substring of a string $S$ is a subsequence of $S$ consists of consecutive elements in $S$. 
We say a string is empty if it does not have any element in it.  An empty string is a subsequence and a substring of any string.
Let $X$ and $Y$ be two strings over an alphabet $\Sigma$.  The longest common subsequence (resp. substring) problem for $X$ and $Y$ is to find the longest string which is a subsequence (resp. substring) of both $X$ and $Y$. Both the longest common subsequence problem and the longest common substring problem
 have applications in different fields. For example, in molecular biology, the lengths of the longest common subsequence and the longest common substring  
can be used to measure the similarity between two biological sequences.
The two problems have been well-investigated in the last several decades.  
More details on the research of the first problem
can be found in \cite{Apo}, \cite {Apo1}, \cite{Bergroth}, \cite{B}, \cite{Cormen}, \cite{Hberg1}, \cite{Hberg}, \cite{Hunt}, \cite{M}, 
\cite{Rick} and references therein and the second problem can be found in \cite{C}, \cite{G}, \cite{W} and references therein.  Motivated by the two problems above, Li, Deka, and Deka \cite{L1} introduced 
the longest common subsequence and substring problem for two strings $X$ and $Y$ which is to find the longest string such that it is a subsequence of $X$ and a substring of $Y$. 
They also proposed an algorithm to solve this problem in \cite{L1}. In this paper, we introduce 
the longest common subsequence and substring problem for multiple strings which is a generalization of the longest common subsequence and substring problem for two stings. Suppose $X_1, X_2, ..., X_s$ and $Y_1, Y_2, ..., Y_t$ are strings over an alphabet $\Sigma$, where $s$ and $t$ are positive integers. 
The ($s$, $t$)-longest common subsequence and substring problem for multiple strings $X_1, X_2, ..., X_s$ and $Y_1, Y_2, ..., Y_t$ is to find the longest string, denoted $Z(X_1, X_2, ..., X_s; Y_1, Y_2, ..., Y_t)$, which is a subsequence of $X_1, X_2, ..., X_s$ and a substring of $Y_1, Y_2, ..., Y_t$. 
If $Z(X_1, X_2, ..., X_s; Y_1, Y_2, ..., Y_t)$ does not exist, we say $Z(X_1, X_2, ..., X_s; Y_1, Y_2, ..., Y_t)$
is an empty string. 
We propose an algorithm to solve the ($s$, $t$)-longest common subsequence and substring problem.  \newline

\noindent {\bf 2.  The Preparations of the Algorithm} \\

Our algorithm is based on several claims to be proved in this section. Before proving the claims, we need some notations as follows.
For a given string $H = h_1 h_2 ... h_k$ over an alphabet $\Sigma$, the size of $H$, denoted $|H|$, is defined as the number of elements in $H$.
The length of an empty string is zero.
The $jth$ suffix of $H$ is the string of $h_j h_{j + 1} ... h_k$, where $1 \leq j \leq k$.
The $ith$ prefix of $H$ is defined as $H[i] = h_1 h_2 ... h_i$, where $1 \leq i \leq k$. Conventionally, $H[0]$ is defined as an empty string.  
  \\

Let 
$X_p = x[p, 1]x[p, 2] ... x[p, m_p]$, where $x[p, a]$ with $p$ is an integer such that $1 \leq p \leq s$ and $1 \leq a \leq m_p$
are elements in an alphabet $\Sigma$, be $s$ strings, and 
$Y_q = y[q, 1]y[q, 2] ... y[q, n_q]$, where  $y[q, b]$ with $q$ is an integer such that $1 \leq q \leq t$ and $1 \leq b \leq n_q$
are elements in the alphabet $\Sigma$, be $t$ strings. We define $Z[i_1, i_2, ..., i_s; j_1, j_2, ..., j_t]$ as a string satisfying the following conditions, 
where $1 \leq i_u \leq m_u$ with $1 \leq u \leq s$ and $1 \leq j_v \leq n_v$ with $1 \leq v \leq t$. \\  

 \hspace*{10mm} ($1.1$) It is a subsequence of $X_1[i_1] = x[1, 1]x[1, 2] ... x[1, i_1]$.   \newline 
 
 \hspace*{10mm} ($1.2$) It is a subsequence of $X_2[i_2] = x[2, 1]x[2, 2] ... x[2, i_2]$.   \newline
 
 \hspace*{54mm}  ............   \newline
 
 \hspace*{10mm} ($1.s$) It is a subsequence of $X_s[i_s] = x[s, 1]x[s, 2] ... x[s, i_s]$.   \newline
 
 \hspace*{10mm} ($2.1$) It is a suffix of $Y_1[j_1] = y[1, 1]y[1, 2] ... y[1, j_1]$.   \newline
 
 \hspace*{10mm} ($2.2$) It is a suffix of $Y_2[j_2] = y[2, 1]y[2, 2] ... y[2, j_2]$.   \newline
 
 \hspace*{54mm}  ............   \newline
 
 \hspace*{10mm} ($2.t$) It is a suffix of $Y_t[j_t] = y[t, 1]y[t, 2] ... y[t, j_t]$.   \newline
 
 \hspace*{10mm} ($3.1$) Under the conditions above, its length is as large as possible. \\ 

\noindent{\bf Claim $1$.} If $y[1, j_1]$, $y[2, j_2]$, ..., $y[t, j_t]$ are not the same, then $Z[i_1, i_2, ..., i_s;$ $j_1, j_2, ..., j_t]$ 
does not exist. Namely, $Z[i_1, i_2, ..., i_s;$ $j_1, j_2, ..., j_t]$ is an empty string.  \\

\noindent {\bf Proof of Claim $1$.} Suppose, to the contrary, that $Z[i_1, i_2, ..., i_s; j_1, j_2, ..., j_t]$ exists. Then 
$Z[i_1, i_2, ..., i_s; j_1, j_2, ..., j_t]$ is not empty. Thus the last
element in it must be equal to each of $y[1, j_1]$, $y[2, j_2]$, ..., $y[t, j_t]$ and $y[1, j_1]$, $y[2, j_2]$, ..., $y[t, j_t]$ are the same,
a contradiction. Therefore $Z[i_1, i_2, ..., i_s; j_1, j_2, ..., j_t]$ does not exist.  \\

Hence the proof of Claim $1$ is complete. \\

\noindent{\bf Claim $2$.} Suppose that $u_1 := y[1, j_1] = y[2, j_2] = \cdots = y[t, j_t]$. If $u_1 = x[1, i_1] = x[2, i_2] = \cdots = x[s, i_s]$,
then $$|Z[i_1, i_2, ..., i_s; j_1, j_2, ..., j_t]| = |Z[i_1 - 1, i_2 - 1, ..., i_s - 1; j_1 - 1, j_2 - 1, ..., j_t - 1]| + 1.$$ 
\noindent {\bf Proof of Claim $2$.} Our proof is divided into two cases. \\

{\noindent \bf Case 1.} $Z[i_1 - 1, i_2 - 1, ..., i_s - 1; j_1 - 1, j_2 - 1, ..., j_t - 1]$ is not empty. \\

 Notice that $Z_{\alpha} := Z[i_1 - 1, i_2 - 1, ..., i_s - 1; j_1 - 1, j_2 - 1, ..., j_t - 1]$ is a subsequence of 
$$X_1[i_1 - 1] := x[1, 1]x[1, 2] ... x[1, i_1 - 1],$$ 
$$X_2[i_2 - 1] := x[2, 1]x[2, 2] ... x[2, i_2 - 1],$$ 
$$............$$ 
$$X_s[i_s - 1] := x[s, 1]x[s, 2] ... x[s, i_s - 1],$$
and a suffix of 
$$Y_1[j_1 - 1] := y[1, 1]y[1, 2] ... y[1, j_1 - 1],$$ 
$$Y_2[j_2 - 1] := y[2, 1]y[2, 2] ... y[2, j_2 - 1],$$ 
$$............$$ 
$$Y_t[j_t - 1] := y[t, 1]y[t, 2] ... y[t, j_t - 1].$$ 
Since $x[1, i_1] = x[2, i_2] = \cdots = x[s, i_s] = u_1 = y[1, j_1] = y[2, j_2] = \cdots = y[t, j_t]$,
we have that $Z_{\alpha}u_1$ is a subsequence of 
$$X_1[i_1] := x[1, 1]x[1, 2] ... x[1, i_1],$$ 
$$X_2[i_2] := x[2, 1]x[2, 2] ... x[2, i_2],$$ 
$$............$$ 
$$X_s[i_s] := x[s, 1]x[s, 2] ... x[s, i_s],$$
and a suffix of 
$$Y_1[j_1] := y[1, 1]y[1, 2] ... y[1, j_1],$$ 
$$Y_2[j_2] := y[2, 1]y[2, 2] ... y[2, j_2],$$ 
$$............$$ 
$$Y_t[j_t] := y[t, 1]y[t, 2] ... y[t, j_t].$$ 
By the definition of $Z[i_1, i_2, ..., i_s; j_1, j_2, ..., j_t]$, we have that 
$$1 \leq |Z[i_1 - 1, i_2 - 1, ..., i_s - 1; j_1 - 1, j_2 - 1, ..., j_t - 1]| + 1$$ 
$$= |Z_{\alpha}u_1| = |Z_{\alpha}| + 1 \leq |Z[i_1, i_2, ..., i_s; j_1, j_2, ..., j_t]|.$$
By the definition of $Z_{\beta} := Z[i_1, i_2, ..., i_s; j_1, j_2, ..., j_t]$, we have that the last element, say $u_2$, in $Z_{\beta}$ must be equal to
 $y[1, j_1]$, $y[2, j_2]$, ..., and $y[t, j_t]$. Thus $u_1 = u_2 = x[1, i_1] = x[2, i_2] = \cdots = x[s, i_s]$. Therefore $Z_{\beta} - u_2$, which is a string obtained from removing $u_2$ from $Z_{\beta}$, is a subsequence of 
$$X_1[i_1 - 1] := x[1, 1]x[1, 2] ... x[1, i_1 - 1],$$ 
$$X_2[i_2 - 1] := x[2, 1]x[2, 2] ... x[2, i_2 - 1],$$ 
$$............$$ 
$$X_s[i_s - 1] := x[s, 1]x[s, 2] ... x[s, i_s - 1],$$
and a suffix of 
$$Y_1[j_1 - 1] := y[1, 1]y[1, 2] ... y[1, j_1 - 1],$$ 
$$Y_2[j_2 - 1] := y[2, 1]y[2, 2] ... y[2, j_2 - 1],$$ 
$$............$$ 
$$Y_t[j_t - 1] := y[t, 1]y[t, 2] ... y[t, j_t - 1].$$ 
By the definition of $Z[i_1 - 1, i_2 - 1, ..., i_s - 1; j_1 - 1, j_2 - 1, ..., j_t - 1]$, we have that 
$$|Z[i_1, i_2, ..., i_s; j_1, j_2, ..., j_t]| - 1 = |Z_{\beta} - u_2|$$
$$= |Z_{\beta}| - 1 \leq |Z[i_1 - 1, i_2 - 1, ..., i_s - 1; j_1 - 1, j_2 - 1, ..., j_t - 1]|.$$
Therefore $$|Z[i_1, i_2, ..., i_s; j_1, j_2, ..., j_t]| = |Z[i_1 - 1, i_2 - 1, ..., i_s - 1; j_1 - 1, j_2 - 1, ..., j_t - 1]| + 1.$$

\noindent {\bf Case 2.} $Z[i_1 - 1, i_2 - 1, ..., i_s - 1; j_1 - 1, j_2 - 1, ..., j_t - 1]$ is empty. \\

Since $u_1$ is a subsequence of 
$$X_1[i_1] := x[1, 1]x[1, 2] ... x[1, i_1],$$ 
$$X_2[i_2] := x[2, 1]x[2, 2] ... x[2, i_2],$$ 
$$............$$ 
$$X_s[i_s] := x[s, 1]x[s, 2] ... x[s, i_s],$$
and a suffix of 
$$Y_1[j_1] := y[1, 1]y[1, 2] ... y[1, j_1],$$ 
$$Y_2[j_2] := y[2, 1]y[2, 2] ... y[2, j_2],$$ 
$$............$$ 
$$Y_t[j_t] := y[t, 1]y[t, 2] ... y[t, j_t].$$
By the definition of $Z[i_1, i_2, ..., i_s; j_1, j_2, ..., j_t]$, we have that 
$$1 = |u_1| \leq |Z[i_1, i_2, ..., i_s; j_1, j_2, ..., j_t]|.$$
 Notice that the proofs for  
$$|Z[i_1, i_2, ..., i_s; j_1, j_2, ..., j_t]| - 1 \leq |Z[i_1 - 1, i_2 - 1, ..., i_s - 1; j_1 - 1, j_2 - 1, ..., j_t - 1]|$$
in the above Case $1$ still hold in this case. We have 
$$0 \leq |Z[i_1, i_2, ..., i_s; j_1, j_2, ..., j_t]| - 1$$ $$\leq |Z[i_1 - 1, i_2 - 1, ..., i_s - 1; j_1 - 1, j_2 - 1, ..., j_t - 1]| = 0.$$
Thus $$|Z[i_1, i_2, ..., i_s; j_1, j_2, ..., j_t]| = 1,$$
$$|Z[i_1 - 1, i_2 - 1, ..., i_s - 1; j_1 - 1, j_2 - 1, ..., j_t - 1]| = 0.$$ 
Therefore $$|Z[i_1, i_2, ..., i_s; j_1, j_2, ..., j_t]| = 
|Z[i_1 - 1, i_2 - 1, ..., i_s - 1; j_1 - 1, j_2 - 1, ..., j_t - 1]| + 1.$$

\hspace*{5mm} Hence the proof of Claim $2$ is complete. \\

\noindent{\bf Claim $3$.} Suppose that $v_1 := y[1, j_1] = y[2, j_2] =, ..., = y[t, j_t]$. If $v_1 \neq x[1, i_1]$, 
$v_1 \neq x[2, i_2]$, ..., and $v_1 \neq x[s, i_s]$,
then $$|Z[i_1, i_2, ..., i_s; j_1, j_2, ..., j_t]| = |Z[i_1 - 1, i_2 - 1, ..., i_s - 1; j_1, j_2, ..., j_t]|.$$ 

\noindent {\bf Proof of Claim $3$.} Our proof is divided into two cases again. \\

\noindent {\bf Case 1.} $Z[i_1 - 1, i_2 - 1, ..., i_s - 1; j_1, j_2, ..., j_t]$ is not empty. \\

Notice that $Z_{\gamma} := Z[i_1 - 1, i_2 - 1, ..., i_s - 1; j_1, j_2, ..., j_t]$ is a subsequence of 
$$X_1[i_1 - 1] := x[1, 1]x[1, 2] ... x[1, i_1 - 1],$$ 
$$X_2[i_2 - 1] := x[2, 1]x[2, 2] ... x[2, i_2 - 1],$$ 
$$............$$ 
$$X_s[i_s - 1] := x[s, 1]x[s, 2] ... x[s, i_s - 1],$$
and a suffix of 
$$Y_1[j_1] := y[1, 1]y[1, 2] ... y[1, j_1],$$ 
$$Y_2[j_2] := y[2, 1]y[2, 2] ... y[2, j_2],$$ 
$$............$$ 
$$Y_t[j_t] := y[t, 1]y[t, 2] ... y[t, j_t].$$ 
Then $Z_{\gamma}$ is a subsequence of 
$$X_1[i_1] := x[1, 1]x[1, 2] ... x[1, i_1],$$ 
$$X_2[i_2] := x[2, 1]x[2, 2] ... x[2, i_2],$$ 
$$............$$ 
$$X_s[i_s] := x[s, 1]x[s, 2] ... x[s, i_s],$$
and a suffix of 
$$Y_1[j_1] := y[1, 1]y[1, 2] ... y[1, j_1],$$ 
$$Y_2[j_2] := y[2, 1]y[2, 2] ... y[2, j_2],$$ 
$$............$$ 
$$Y_t[j_t] := y[t, 1]y[t, 2] ... y[t, j_t].$$ 
By the definition of $Z[i_1, i_2, ..., i_s; j_1, j_2, ..., j_t]$, we have that
 $$|Z[i_1 - 1, i_2 - 1, ..., i_s - 1; j_1, j_2, ..., j_t]| = |Z_{\gamma}| \leq |Z[i_1, i_2, ..., i_s; j_1, j_2, ..., j_t]|.$$
 Since $Z[i_1 - 1, i_2 - 1, ..., i_s - 1; j_1, j_2, ..., j_t]$ is not empty, 
 $Z_{\delta} := Z[i_1, i_2, ..., i_s;$ $j_1, j_2, ..., j_t]$ is not empty. Thus the last element, say $v_2$, in $Z_{\delta}$ must be equal to
 $y[1, j_1], y[2, j_2], ...,$ $y[t, j_t]$. Thus $v_1 = v_2 = y[1, j_1] = y[2, j_2] = \cdots = y[t, j_t]$. 
 Since $v_2 = v_1 \neq x[1, i_1]$, 
 $v_2 = v_1 \neq x[2, i_2]$, ..., $v_2 = v_1 \neq x[s, i_s]$, we have that $Z_{\delta}$ is a subsequence of 
$$X_1[i_1 - 1] := x[1, 1]x[1, 2] ... x[1, i_1 - 1],$$ 
$$X_2[i_2 - 1] := x[2, 1]x[2, 2] ... x[2, i_2 - 1],$$ 
$$............$$ 
$$X_s[i_s - 1] := x[s, 1]x[s, 2] ... x[s, i_s - 1],$$
and a suffix of 
$$Y_1[j_1] := y[1, 1]y[1, 2] ... y[1, j_1],$$ 
$$Y_2[j_2] := y[2, 1]y[2, 2] ... y[2, j_2],$$ 
$$............$$ 
$$Y_t[j_t] := y[t, 1]y[t, 2] ... y[t, j_t].$$ 
By the definition of $Z[i_1 - 1, i_2 - 2, ..., i_s - 1; j_1, j_2, ..., j_t]$, we have that 
$$|Z[i_1, i_2, ..., i_s; j_1, j_2, ..., j_t]| \leq |Z[i_1 - 1, i_2 - 1, ..., i_s - 1; j_1, j_2, ..., j_t]|.$$
Therefore $$|Z[i_1, i_2, ..., i_s; j_1, j_2, ..., j_t]| = |Z[i_1 - 1, i_2 - 1, ..., i_s - 1; j_1, j_2, ..., j_t]|.$$
 
 \noindent {\bf Case 2.} $Z[i_1, i_2, ..., i_s; j_1, j_2, ..., j_t]$ is empty. \\
 
 Our assertion is that $Z[i_1, i_2, ..., i_s; j_1, j_2, ..., j_t]$ must be empty. Suppose, to the contrary, that 
 $Z_{\nu} := Z[i_1, i_2, ..., i_s; j_1, j_2, ..., j_t]$ is not empty. Then the last element, say $v_3$, in $Z_{\nu}$ must be equal to
 $y[1, j_1], y[2, j_2], ..., y[t, j_t]$. Thus $v_1 = v_3 = y[1, j_1] = y[2, j_2] = \cdots = y[t, j_t]$. 
 Since $v_3 = v_1 \neq x[1, i_1]$, 
 $v_3 = v_1 \neq x[2, i_2]$, ..., $v_3 = v_1 \neq x[s, i_s]$, we have that $Z_{\nu}$ is a subsequence of 
$$X_1[i_1 - 1] := x[1, 1]x[1, 2] ... x[1, i_1 - 1],$$ 
$$X_2[i_2 - 1] := x[2, 1]x[2, 2] ... x[2, i_2 - 1],$$ 
$$............$$ 
$$X_s[i_s - 1] := x[s, 1]x[s, 2] ... x[s, i_s - 1],$$
and a suffix of 
$$Y_1[j_1] := y[1, 1]y[1, 2] ... y[1, j_1],$$ 
$$Y_2[j_2] := y[2, 1]y[2, 2] ... y[2, j_2],$$ 
$$............$$ 
$$Y_t[j_t] := y[t, 1]y[t, 2] ... y[t, j_t].$$ 
By the definition of $Z[i_1 - 1, i_2 - 1, ..., i_s - 1; j_1, j_2, ..., j_t]$, we have
 $$1 \leq |Z[i_1, i_2, ..., i_s; j_1, j_2, ..., j_t]| \leq |Z[i_1 - 1, i_2 - 1, ..., i_s - 1; j_1, j_2, ..., j_t]| = 0,$$
 a contradiction.
 Thus $Z[i_1, i_2, ..., i_s; j_1, j_2, ..., j_t]$ is empty and
 $$|Z[i_1, i_2, ..., i_s; j_1, j_2, ..., j_t]| = |Z[i_1 - 1, i_2 - 1, ..., i_s - 1; j_1, j_2, ..., j_t]| = 0.$$ 

\hspace*{5mm} Hence the proof of Claim $3$ is complete. \\

\noindent{\bf Claim $4$.} Suppose that $w_1 := y[1, j_1] = y[2, j_2] = \cdots = y[t, j_t]$. Assume that $w_1$ is not equal to 
exactly $r$ elements in the set $L: = \{x[1, i_1], x[2, i_2], ..., x[s, i_s]\}$, where $1 \leq r \leq (s - 1)$. 
Without loss of generality, we assume that $w_1$ is not equal to exactly the first $r$ elements in $L$. Namely, 
$w_1 \neq x[1, i_1]$, $w_1 \neq x[2, i_2]$, ...,  $w_1 \neq x[r, i_r]$, and $w_1 = x[r + 1, i_{r + 1}] = x[r + 2, i_{r + 2}] = \cdots 
= x[s, i_s]$.
Then $$|Z[i_1, i_2, ..., i_s; j_1, j_2, ..., j_t]|$$
$$= |Z[i_1 - 1, i_2 - 1, ...,i_r - 1, i_{r + 1}, i_{r + 2}, ..., i_s; j_1, j_2, ..., j_t]|.$$ 

\noindent {\bf Proof of Claim $4$.} Our proof is divided into two cases. \\

\noindent {\bf Case 1.} $Z[i_1 - 1, i_2 - 1, ...,i_r - 1, i_{r + 1}, i_{r + 2}, ..., i_s; j_1, j_2, ..., j_t]$ is not empty. \\

Notice that $Z_{\epsilon} := Z[i_1 - 1, i_2 - 1, ..., i_r - 1, i_{r + 1}, i_{r + 2}, ..., i_s;$ 
$j_1, j_2, ..., j_t]$ is a subsequence of 
$$X_1[i_1 - 1] := x[1, 1]x[1, 2] ... x[1, i_1 - 1],$$ 
$$X_2[i_2 - 1] := x[2, 1]x[2, 2] ... x[2, i_2 - 1],$$ 
$$............$$ 
$$X_r[i_r - 1] := x[r, 1]x[r, 2] ... x[r, i_r - 1],$$ 
$$X_{r + 1}[i_{r + 1} - 1] := x[r + 1, 1]x[r + 1, 2] ... x[r + 1, i_{r + 1}],$$
$$X_{r + 2}[i_{r + 2} - 1] := x[r + 2, 1]x[r + 2, 2] ... x[r + 2, i_{r + 2}],$$
$$............$$  
$$X_s[i_s - 1] := x[s, 1]x[s, 2] ... x[s, i_s],$$
and a suffix of 
$$Y_1[j_1] := y[1, 1]y[1, 2] ... y[1, j_1],$$ 
$$Y_2[j_2] := y[2, 1]y[2, 2] ... y[2, j_2],$$ 
$$............$$ 
$$Y_t[j_t] := y[t, 1]y[t, 2] ... y[t, j_t].$$ 
Then $Z_{\gamma}$ is a subsequence of 
$$X_1[i_1] := x[1, 1]x[1, 2] ... x[1, i_1],$$ 
$$X_2[i_2] := x[2, 1]x[2, 2] ... x[2, i_2],$$ 
$$............$$ 
$$X_s[i_s] := x[s, 1]x[s, 2] ... x[s, i_s],$$
and a suffix of 
$$Y_1[j_1] := y[1, 1]y[1, 2] ... y[1, j_1],$$ 
$$Y_2[j_2] := y[2, 1]y[2, 2] ... y[2, j_2],$$ 
$$............$$ 
$$Y_t[j_t] := y[t, 1]y[t, 2] ... y[t, j_t].$$ 
By the definition of $Z[i_1, i_2, ..., i_s; j_1, j_2, ..., j_t]$, we have that
 $$|Z[i_1 - 1, i_2 - 1, ..., i_r - 1, i_{r + 1}, i_{r + 2}, ..., i_s; j_1, j_2, ..., j_t]| = |Z_{\epsilon}|$$
 $$\leq |Z[i_1, i_2, ..., i_s; j_1, j_2, ..., j_t]|.$$
 Since $Z[i_1 - 1, i_2 - 1, ...,i_r - 1, i_{r + 1}, i_{r + 2}, ..., i_s; j_1, j_2, ..., j_t]$ is not empty, 
 $Z_{\mu} := Z[i_1, i_2, ..., i_s; j_1, j_2, ..., j_t]$ is not empty. 
Thus the last element, say $w_2$, in $Z_{\mu}$ must be equal to
 $y[1, j_1]$, $y[2, j_2]$, ..., and $y[t, j_t]$. Thus $w_1 = w_2 \neq x[1, i_1]$, $w_1 = w_2 \neq x[2, i_2]$, ...,  $w_1 = w_2 \neq x[r, i_r]$,  
 and $w_1 = w_2 = x[r + 1, i_{r + 1}] = x[r + 2, i_{r + 2}] = \cdots = x[s, i_s]$. 
 Therefore $Z_{\mu}$ is a subsequence of 
$$X_1[i_1 - 1] := x[1, 1]x[1, 2] ... x[1, i_1 - 1],$$ 
$$X_2[i_2 - 1] := x[2, 1]x[2, 2] ... x[2, i_2 - 1],$$ 
$$............$$ 
$$X_r[i_r - 1] := x[r, 1]x[r, 2] ... x[r, i_r - 1],$$ 
$$X_{r + 1}[i_{r + 1} - 1] := x[r + 1, 1]x[r + 1, 2] ... x[r + 1, i_{r + 1}],$$
$$X_{r + 2}[i_{r + 2} - 1] := x[r + 2, 1]x[r + 2, 2] ... x[r + 2, i_{r + 2}],$$
$$............$$  
$$X_s[i_s - 1] := x[s, 1]x[s, 2] ... x[s, i_s],$$
and a suffix of 
$$Y_1[j_1] := y[1, 1]y[1, 2] ... y[1, j_1],$$ 
$$Y_2[j_2] := y[2, 1]y[2, 2] ... y[2, j_2],$$ 
$$............$$ 
$$Y_t[j_t] := y[t, 1]y[t, 2] ... y[t, j_t].$$
 By the definition of $Z[i_1 - 1, i_2 - 1, ...,i_r - 1, i_{r + 1}, i_{r + 2}, ..., i_s; j_1, j_2, ..., j_t]$, we have that 
$$|Z[i_1, i_2, ..., i_s; j_1, j_2, ..., j_t]|$$ $$\leq |Z[i_1 - 1, i_2 - 1, ...,i_r - 1, i_{r + 1}, i_{r + 2}, ..., i_s; j_1, j_2, ..., j_t]|.$$
Therefore $$|Z[i_1, i_2, ..., i_s; j_1, j_2, ..., j_t]|$$ $$= |Z[i_1 - 1, i_2 - 1, ...,i_r - 1, i_{r + 1}, i_{r + 2}, ..., i_s; j_1, j_2, ..., j_t]|.$$

\noindent {\bf Case 2.} $Z[i_1 - 1, i_2 - 1, ...,i_r - 1, i_{r + 1}, i_{r + 2}, ..., i_s; j_1, j_2, ..., j_t]$ is empty. \\

 Our assertion is that $Z[i_1, i_2, ..., i_s; j_1, j_2, ..., j_t]$ must be empty. Suppose, to the contrary, that 
 $Z_{\rho} := Z[i_1, i_2, ..., i_s; j_1, j_2, ..., j_t]$ is not empty. Then the last element, say $w_3$, in $Z_{\rho}$ must be equal to
 $y[1, j_1], y[2, j_2], ..., y[t, j_t]$.  Thus $w_1 = w_3 \neq x[1, i_1]$, $w_1 = w_3 \neq x[2, i_2]$, ...,  $w_1 = w_3 \neq x[r, i_r]$,  
 and $w_1 = w_3 = x[r + 1, i_{r + 1}] = x[r + 2, i_{r + 2}] = \cdots = x[s, i_s]$.  Therefore $Z_{\rho}$ is a subsequence of 
$$X_1[i_1 - 1] := x[1, 1]x[1, 2] ... x[1, i_1 - 1],$$ 
$$X_2[i_2 - 1] := x[2, 1]x[2, 2] ... x[2, i_2 - 1],$$ 
$$............$$ 
$$X_r[i_r - 1] := x[r, 1]x[r, 2] ... x[r, i_r - 1],$$ 
$$X_{r + 1}[i_{r + 1} - 1] := x[r + 1, 1]x[r + 1, 2] ... x[r + 1, i_{r + 1}],$$
$$X_{r + 2}[i_{r + 2} - 1] := x[r + 2, 1]x[r + 2, 2] ... x[r + 2, i_{r + 2}],$$
$$............$$  
$$X_s[i_s - 1] := x[s, 1]x[s, 2] ... x[s, i_s],$$
and a suffix of 
$$Y_1[j_1] := y[1, 1]y[1, 2] ... y[1, j_1],$$ 
$$Y_2[j_2] := y[2, 1]y[2, 2] ... y[2, j_2],$$ 
$$............$$ 
$$Y_t[j_t] := y[t, 1]y[t, 2] ... y[t, j_t].$$
By the definition of $Z[i_1 - 1, i_2 - 1, ...,i_r - 1, i_{r + 1}, i_{r + 2}, ..., i_s; j_1, j_2, ..., j_t]$, we have that 
$$1 \leq |Z[i_1, i_2, ..., i_s; j_1, j_2, ..., j_t]|$$ 
$$\leq |Z[i_1 - 1, i_2 - 1, ...,i_r - 1, i_{r + 1}, i_{r + 2}, ..., i_s; j_1, j_2, ..., j_t]| = 0,$$
a contradiction. Thus $Z[i_1, i_2, ..., i_s; j_1, j_2, ..., j_t]$ is empty and 
$$|Z[i_1, i_2, ..., i_s; j_1, j_2, ..., j_t]|$$
$$= |Z[i_1 - 1, i_2 - 1, ...,i_r - 1, i_{r + 1}, i_{r + 2}, ..., i_s; j_1, j_2, ..., j_t]| = 0.$$

\hspace*{5mm} Hence the proof of Claim $4$ is complete. \\

\noindent{\bf Remark $1$.} The general form of Claim $4$ is as follows. \newline
\noindent{\bf Claim $4'$.} Suppose that $w_1 := Y[1, j_1] = Y[2, j_2] =, ..., = Y[t, j_t]$. 
If $w_1 \neq x[\pi(1), i_{\pi(1)}]$, $w_1 = x[\pi(2), i_{\pi(2)}]$, ...,  $w_1 \neq x[\pi(r), i_{\pi(r)}]$, 
where $\pi(1), \pi(2), ...,  \pi(r)$ are integers such that $1 \leq \pi(1) < \pi(2)< \cdots < \pi(r) \leq s$, and
for any $e \in \{\, 1, 2, ..., s \,\} - \{\, \pi(1), \pi(2), ..., \pi(r) \,\}$, $w_1 = x[e, i_{e}]$, then 
$$|Z[i_1, i_2, ..., i_s; j_1, j_2, ..., j_t]|$$ 
$$= |Z[i_1, ..., i_{\pi(1) - 1}, i_{\pi(1)} - 1, i_{\pi(1) + 1}, ..., i_{\pi(2) - 1}, i_{\pi(2)} - 1, i_{\pi(2) + 1}, ..., $$
$$i_{\pi(r) - 1}, i_{\pi(r)} - 1, i_{\pi(r) + 1}, ..., i_s; j_1, j_2, ..., j_t]|.$$
\noindent{\bf Remark $2$.} Suppose that $w_1 := y[1, j_1] = y[2, j_2] =, ..., = y[t, j_t]$. We need to 
follow Claim $2$, Claim $3$, and Claim $4$ to compute $|Z[i_1, i_2, ..., i_s; j_1, j_2, ..., j_t]|$.
The largest number of formats we can encounter is  
$$C(s, 0) + C(s, 1) + C(s, 2) + \cdots + C(s, s) = 2^s,$$ 
where $C(s, a)$ denotes the number of $a$-element subsets of a set of size $s$, where $a$ is an integer such that $0 \leq a \leq s$.   \\

\noindent{\bf Claim 5.} Let $H = h_1 h_2 ... h_b$  be a longest string which is a subsequence of $X_1$, $X_2$, ...,  $X_s$,
and a substring of $Y_1$, $Y_2$, ..., $Y_n$. Then 
$$b = \max \{\, |Z[i_1, i_2, ..., i_s; j_1, j_2, ..., j_t]| : 1 \leq i_1 \leq m_1, 1 \leq i_2 \leq m_2, ..., $$ $$1 \leq i_s \leq m_s,
1 \leq j_1 \leq n_1, 1 \leq j_2 \leq n_2, ...,  1 \leq j_t \leq n_t \,\}.$$ 
\noindent {\bf Proof of Claim 5.} For any $i_1$, $i_2$, ..., $i_s$ with $1 \leq i_1 \leq m_1$, $1 \leq i_2 \leq m_2$, ..., $1 \leq i_s \leq m_s$, 
and any $j_1$, $j_2$, ..., $j_t$ with $1 \leq j_1 \leq n_1$, $1 \leq j_2 \leq n_2$, ..., $1 \leq j_t \leq n_t$, 
 we, from the definition of $Z[i_1, i_2, ..., i_s; j_1, j_2, ..., j_t]$, have that $Z[i_1, i_2, ..., i_s; j_1, j_2, ..., j_t]$ is a subsequence of 
 $X_1$, $X_2$, ..., $X_s$,
and a substring of $Y_1$, $Y_2$, ..., $Y_n$.
 By the definition of $H$, we have that 
 $$|Z[i_1, i_2, ..., i_s; j_1, j_2, ..., j_t]| \leq |H| = b.$$
 Thus $$\max \{\, |Z[i_1, i_2, ..., i_s; j_1, j_2, ..., j_t]| : 1 \leq i_1 \leq m_1, 1 \leq i_2 \leq m_2, ..., $$ $$1 \leq i_s \leq m_s,
1 \leq j_1 \leq n_1, 1 \leq j_2 \leq n_2, ...,  1 \leq j_t \leq n_t \,\} \leq b.$$ 
Since $H = h_1 h_2 ... h_b$  is a longest string which is a subsequence of $X_1$, $X_2$, ..., $X_s$,
and a substring of $Y_1$, $Y_2$, ..., $Y_n$, there exits indices $i_1$, $i_2$, ..., $i_s$ and indices
$j_1$, $j_2$, ..., $j_t$ such that $h_b = x[1, i_1]$, $h_b = x[2, i_2]$, ..., $h_b = x[s, i_s]$, and $h_b = y[1, j_1]$, $h_b = y[2, j_2]$, ...,
$h_b = y[t, j_t]$. Thus
$H = h_1 h_2 ... h_b$ is a subsequence of $X_1[i_1]$, $X_2[i_2]$, ..., $X_s[i_s]$ and a suffix of 
$Y_1[j_1]$, $Y_2[j_2]$, ..., $Y_t[j_t]$. From the definition of $Z[i_1, i_2, ..., i_s; j_1, j_2, ..., j_t]$, we have that 
$$b \leq |Z[i_1, i_2, ..., i_s; j_1, j_2, ..., j_t]|$$ 
$$\leq \max \{\, |Z[i_1, i_2, ..., i_s; j_1, j_2, ..., j_t]| : 1 \leq i_1 \leq m_1, 1 \leq i_2 \leq m_2, ..., $$ $$1 \leq i_s \leq m_s,
1 \leq j_1 \leq n_1, 1 \leq j_2 \leq n_2, ...,  1 \leq j_t \leq n_t \,\}.$$
Therefore $$b = \max \{\, |Z[i_1, i_2, ..., i_s; j_1, j_2, ..., j_t]| : 1 \leq i_1 \leq m_1, 1 \leq i_2 \leq m_2, ..., $$ 
$$1 \leq i_s \leq m_s, 1 \leq j_1 \leq n_1, 1 \leq j_2 \leq n_2, ...,  1 \leq j_t \leq n_t \,\}.$$
\hspace*{5mm} Hence the proof of Claim $5$ is complete. \\

\noindent {\bf 3. An Algorithm for the ($s$, $t$)-Longest Common Subsequence and Substring Problem} \newline

Based on Claims $1$-$5$ in Section $2$, we can design an algorithm for the ($s$, $t$)-longest common subsequence and substring problem. 
Once again, we assume that $X_p = x[p, 1]x[p, 2] ... x[p, m_p]$, 
where $x[p, a]$ with $p$ is an integer such that $1 \leq p \leq s$ and $1 \leq a \leq m_p$ are elements in the alphabet $\Sigma$,
are $s$ strings, and 
$Y_q = y[q, 1]y[q, 2] ... y[q, n_q]$, where  $y[q, b]$ with $q$ is an integer such that $1 \leq q \leq t$ and $1 \leq b \leq n_q$
are elements in the alphabet $\Sigma$, are $t$ strings. In the following Algorithm $A$, 
$W$ is an $(m_1 + 1) \times (m_2 + 1) \times \cdots \times (m_s + 1) \times (n_1 + 1) \times (n_2 + 1) \times \cdots \times(n_t + 1)$ array and
the cells $W(i_1, i_2, ..., i_s, j_1, j_2, ..., j_t)$, where $1 \leq i_1 \leq m_1$, $1 \leq i_2 \leq m_2$, ..., $1 \leq i_s \leq m_s$,
 and $1 \leq j_1 \leq n_1$, $1 \leq j_2 \leq n_2$, ..., $1 \leq j_t \leq n_t$ store the lengths of strings such that each of them satisfies the following conditions. \newline  
 
\hspace*{10mm} ($1.1$) It is a subsequence of $X_1[i_1] = x[1, 1]x[1, 2] ... x[1, i_1]$.   \newline 
 
 \hspace*{10mm} ($1.2$) It is a subsequence of $X_2[i_2] = x[2, 1]x[2, 2] ... x[2, i_2]$.   \newline
 
 \hspace*{54mm}  ............   \newline
 
 \hspace*{10mm} ($1.s$) It is a subsequence of $X_s[i_s] = x[s, 1]x[s, 2] ... x[s, i_s]$.   \newline
 
 \hspace*{10mm} ($2.1$) It is a suffix of $Y_1[j_1] = y[1, 1]y[1, 2] ... y[1, j_1]$.   \newline
 
 \hspace*{10mm} ($2.2$) It is a suffix of $Y_2[j_2] = y[2, 1]y[2, 2] ... y[2, j_2]$.   \newline
 
 \hspace*{54mm}  ............   \newline
 
 \hspace*{10mm} ($2.t$) It is a suffix of $Y_t[j_t] = y[t, 1]y[t, 2] ... y[t, j_t]$.   \newline
 
 \hspace*{10mm} ($3.1$) Under the conditions above, its length is as large as possible. \\

\noindent {\bf $ALG \, A(X_1, X_2, ..., X_m, Y_1, Y_2, ..., Y_n, m, n, W)$} \\
$1.$ Initialization: $W(i_1, i_2, ..., i_s, j_1, j_2, ..., j_t) \leftarrow 0$, where $0 \leq i_1 \leq m_1$,  \\
\hspace*{31mm} $i_2 = 0$, ..., $i_s = 0$, $j_1 = 0$, $j_2 = 0$, ..., $j_t = 0$.  \\ 

\hspace*{25mm} $W(i_1, i_2, ..., i_s, j_1, j_2, ..., j_t) \leftarrow 0$, where $i_1 = 0$, $0 \leq i_2 \leq m_2$, \\ 
\hspace*{32mm} $i_3 = 0$, ..., $i_s = 0$, $j_1 = 0$, $j_2 = 0$, ..., $j_t = 0$.  \\ 

\hspace*{76mm} ............ \\ 

\hspace*{25mm} $W(i_1, i_2, ..., i_s, j_1, j_2, ..., j_t) \leftarrow 0$, where $i_1 = 0$, ..., $i_{s - 1} = 0$, 
\hspace*{32mm} $0 \leq i_s \leq m_s$, $j_1 = 0$, $j_2 = 0$, ..., $j_t = 0$.  \\ 

\hspace*{25mm} $W(i_1, i_2, ..., i_s, j_1, j_2, ..., j_t) \leftarrow 0$, where $i_1 = 0$, ..., $i_s = 0$,
\hspace*{32mm} $0 \leq j_1 \leq n_1$, $j_2 = 0$, $j_3 = 0$, ..., $j_t = 0$.  \\ 

\hspace*{25mm} $W(i_1, i_2, ..., i_s, j_1, j_2, ..., j_t) \leftarrow 0$, where $i_1 = 0$, ..., $i_s = 0$, 
\hspace*{31mm} $j_1 = 0$, $0 \leq j_2 \leq n_2$, $j_3 = 0$, ..., $j_t = 0$.  \\ 

\hspace*{76mm} ............ \\ 

\hspace*{25mm} $W(i_1, i_2, ..., i_s, j_1, j_2, ..., j_t) \leftarrow 0$, where $i_1 = 0$, ..., $i_{s} = 0$,
\hspace*{32mm} $j_1 = 0$, $j_2 = 0$, ..., $j_{t - 1} = 0$, $0 \leq j_t \leq n_t$.  \\ 

 \hspace*{26mm}     $maxLength = 0$  \\ 
 
\hspace*{26mm}      $lastIndexOnY1 = n_1$ \\ 

\noindent $2.1.$ {\bf for} $\theta_1 \leftarrow 1$ {\bf to} $m_1$  \newline
\noindent $2.2.$ \hspace*{2mm}{\bf for} $\theta_2 \leftarrow 1$ {\bf to} $m_2$  \newline
\hspace*{14mm} ............ \newline
\noindent $2.s.$ \hspace*{4mm}{\bf for} $\theta_s \leftarrow 1$ {\bf to} $m_s$  \newline
 \noindent$3.1.$ \hspace*{5mm} {\bf for} $\tau_1 \leftarrow 1$ {\bf to} $n_1$  \newline
\noindent $3.2.$ \hspace*{8mm}{\bf for} $\tau_2 \leftarrow 1$ {\bf to} $n_2$  \newline
\hspace*{22mm} ............ \newline
\noindent $3.t.$ \hspace*{11mm}{\bf for} $\tau_t \leftarrow 1$ {\bf to} $n_t$  \newline
\hspace*{20mm} {\bf if}  $y[1, \tau_1]$, $y[2, \tau_2]$, ..., $y[t, \tau_t]$ are not the same \\
    \hspace*{27mm}         $W(\theta_1, \theta_2, ..., \theta_s, \tau_1, \tau_2, ..., \tau_t) \leftarrow 0$ \newline
\hspace*{20mm} {\bf else}  \newline
  \hspace*{28mm} Set $\sigma := y[1, \tau_1] = y[2, \tau_2] = x[t, \tau_t]$  \newline               
\hspace*{28mm} {\bf if} $\sigma = x[1, \theta_1] = x[2, \theta_2] = \cdots = x[s, \theta_s]$ \newline
\hspace*{35mm} $W(i_1, i_2, ..., i_s, j_1, j_2, ..., j_t) \leftarrow W(i_1 - 1, i_2 - 1, ..., $ \newline 
\hspace*{36mm} $i_s - 1, j_1 - 1, j_2 - 1, ..., j_t - 1) + 1$ \newline
\hspace*{28mm} {\bf else if} $\sigma \neq x[1, \theta_1]$,  $\sigma \neq x[2, \theta_2]$, ..., $\sigma \neq x[s, \theta_s]$ \newline
\hspace*{36mm} $W(i_1, i_2, ..., i_s, j_1, j_2, ..., j_t) \leftarrow W(i_1 - 1, i_2 - 1, ..., $ \newline 
\hspace*{37mm} $i_s - 1, j_1, j_2, ..., j_t)$ \newline
\hspace*{29mm} {\bf else} $\sigma \neq x[\pi(1), i_{\pi(1)}]$, $\sigma \neq x[\pi(2), i_{\pi(2)}]$, ...,  $\sigma \neq x[\pi(r), i_{\pi(r)}]$, 
\hspace*{36mm} where $1 \leq \pi(1) < \pi(2)< \cdots < \pi(r) \leq s$, $1 \leq r \leq$ 
\hspace*{37mm} $(s - 1)$, and for any $e \in \{1, 2, ..., s\} - \{\pi(1), \pi(2), ..., \pi(r)\}$, \newline
\hspace*{37mm} $\sigma = x[e, i_{e}]$, \newline
\hspace*{37mm} $W(i_1, i_2, ..., i_s, j_1, j_2, ..., j_t) \leftarrow W(i_1, ..., i_{\pi(1) - 1}, i_{\pi(1)} - 1, $ \newline 
\hspace*{38mm} $i_{\pi(1) + 1}, ..., i_{\pi(2) - 1}, i_{\pi(2)} - 1, i_{\pi(2) + 1}, ..., i_{\pi(r) - 1}, i_{\pi(r)} - 1,$ \newline
\hspace*{38mm} $i_{\pi(r) + 1}, ..., i_s; j_1, j_2, ..., j_t)$ \newline
\hspace*{24mm} {\bf if} $W(i_1, i_2, ..., i_s; j_1, j_2, ..., j_t) > maxLength$ \\
\hspace*{32mm}  $maxLength = W(i_1, i_2, ..., i_s; j_1, j_2, ..., j_t)$ \\
\hspace*{33mm}  $lastIndexOnY1 = j_1$ \\ \\
\noindent $4.$ {\bf return} $A \,\, substring \,\, of \, Y_1 \,\, between \,\, (lastIndexOnY1 \, - \, maxLength)$ \newline
\hspace*{19mm} $and \,\, lastIndexOnY1$ \\

Because of Claims $1$-$5$ in Section $2$, we have that Algorithm $A$ is correct. Clearly, the time complexity of 
Algorithm $A$ is $O(m_1 \, m_2\, \cdots \, m_s \, n_1 \, n_2, \, \cdots \, n_t)$ and the space complexity of Algorithm $A$ is also 
$O(m_1 \, m_2\, \cdots \, m_s \, n_1 \, n_2, \, \cdots \, n_t)$.  \\




\end{document}